\documentclass[lettersize,journal]{IEEEtran}
\usepackage{amsmath,amsfonts}
\usepackage{algorithmic}
\usepackage{algorithm}
\usepackage{array}
\usepackage[caption=false,font=normalsize,labelfont=sf,textfont=sf]{subfig}
\usepackage{textcomp}
\usepackage{stfloats}
\usepackage{url}
\usepackage{verbatim}
\usepackage{graphicx}
\usepackage{cite}

\usepackage{hhline}
\usepackage{colortbl}
\usepackage{threeparttable}
\usepackage{booktabs}
\usepackage{multirow}
\usepackage{multicol}   
\usepackage{array}

\begin{document}

\title{Large Language Model-driven Security Assistant for Internet of Things via Chain-of-Thought}
\author{Mingfei Zeng,
        Ming Xie,
        Xixi Zheng,
        Chunhai Li,
        Chuan Zhang,
        Liehuang Zhu

\thanks{Mingfei Zeng and Ming Xie are with Guangxi Power Grid Co., Ltd., China. Email: zeng\_mf.xt@gx.csg.cn, wolfgangtse@qq.com.}

\thanks{Xixi Zheng, Chuan Zhang, and Liehuang Zhu are with School of Cyberspace Science and Technology, Beijing Institute of Technology. Email: bit-zhengxixi@bit.edu.cn, chuanz@bit.edu.cn, liehuangz@bit.edu.cna.}

\thanks{Chunhai Li is with Guangxi Engineering Research Center of Industrial Internet Security and Blockchain, Guilin University of Electronic Technology. Email: chunhaili@guet.edu.cn.}


\thanks{This work is supported by the National Natural Science Foundation of China (Grant No. 62472032), the Open Foundation of Key Laboratory of Cyberspace Security, Ministry of Education of China (Grant No.KLCS20240209), and the Young Elite Scientists Sponsorship Program by CAST (Grant No. 2023QNRC001).}
}

\maketitle

\begin{abstract}
The rapid development of Internet of Things (IoT) technology has transformed people's way of life and has a profound impact on both production and daily activities. However, with the rapid advancement of IoT technology, the security of IoT devices has become an unavoidable issue in both research and applications. Although some efforts have been made to detect or mitigate IoT security vulnerabilities, they often struggle to adapt to the complexity of IoT environments, especially when dealing with dynamic security scenarios. How to automatically, efficiently, and accurately understand these vulnerabilities remains a challenge. To address this, we propose an IoT security assistant driven by Large Language Model (LLM), which enhances the LLM's understanding of IoT security vulnerabilities and related threats. The aim of the ICoT method we propose is to enable the LLM to understand security issues by breaking down the various dimensions of security vulnerabilities and generating responses tailored to the user's specific needs and expertise level. By incorporating ICoT, LLM can gradually analyze and reason through complex security scenarios, resulting in more accurate, in-depth, and personalized security recommendations and solutions. Experimental results show that, compared to methods relying solely on LLM, our proposed LLM-driven IoT security assistant significantly improves the understanding of IoT security issues through the ICoT approach and provides personalized solutions based on the user's identity, demonstrating higher accuracy and reliability.
\end{abstract}


\section{Introduction}
\IEEEPARstart{T}{he} rapid development of the Internet of Things (IoT) has profoundly transformed both industrial and everyday life by connecting a wide array of devices and systems, significantly enhancing operational efficiency, convenience, and overall productivity. IoT technologies enable devices to communicate in real time, driving innovations in various sectors such as healthcare, smart homes, transportation, and smart grids~\cite{al2024role,zhang2024vulnerability}. Hundreds of billions of IoT devices have already been integrated into home and industrial environments, fundamentally changing the way people live~\cite{dong2025chatiot}. However, alongside these advancements, the rapid proliferation of IoT devices also introduces significant security challenges that cannot be overlooked.

As the number of devices interconnected increases, there is a dramatic rise in cyberattacks targeting IoT devices~\cite{mutleg2024comprehensive}. These attacks exploit vulnerabilities in both the hardware and software of IoT devices, often with the intent of stealing sensitive data, disrupting services, or even gaining unauthorized control over critical systems. Although considerable progress has been made in IoT security research, the complexity of IoT environments characterized by heterogeneous devices, diverse communication protocols, and varying levels of security standards presents ongoing challenges in identifying, understanding, and mitigating vulnerabilities~\cite{mahadik2024heterogeneous}. In such a dynamic and evolving landscape, developing robust solutions to accurately detect and address IoT security issues remains a pressing concern for both researchers and practitioners alike.

Several studies have attempted to detect and analyze IoT vulnerabilities in a traditional way~\cite{shoshitaishvili2015firmalice, corteggiani2018inception, zheng2019firm}, and with the development of artificial intelligence, some research has leveraged machine learning and deep learning to bring new insights into IoT security~\cite{al2020survey, hussain2020machine, shafiq2020corrauc, vasan2020mthael, chaganti2022deep}. However, existing methods often rely on predefined rules or static threat models, which are difficult to adapt to the constantly evolving and diverse nature of IoT environments. Additionally, the complexity of IoT systems, involving heterogeneous devices, protocols, and data streams, makes it challenging to provide comprehensive and actionable security insights, particularly when assisting users with different levels of knowledge.

To address these challenges, we propose a LLM-driven Security Assistant that provides personalized and practical vulnerability analysis and solutions for different types of users. Specifically, we introduce the IoT Chain-of-Thought (ICoT), which aims to enhance the LLM’s understanding of IoT security vulnerabilities and threats by breaking down the characteristics of security issues and providing personalized security advice based on specific user needs and expertise levels. Extensive experiments have confirmed that ICoT significantly improves the ability to detect and resolve security vulnerabilities in IoT systems, offering a novel solution for strengthening IoT security in an increasingly interconnected world. To summarize, our main contributions are as follows:

\begin{itemize}
    \item We introduce an innovative IoT security assistant driven by LLMs that improves the understanding of IoT security vulnerabilities and threats, offering personalized responses tailored to users with diverse professional backgrounds.
    \item Our proposed ICoT method breaks down the characteristics of security vulnerabilities, enabling the LLM to generate more accurate security recommendations for IoT environments, without the need for task-specific fine-tuning or domain-specific datasets, making it widely applicable to a variety of IoT security scenarios.
    \item Extensive experimental evaluations demonstrate that ICoT significantly improves the identification and mitigation of IoT security vulnerabilities, outperforming traditional methods in both accuracy and reliability.
\end{itemize}

\section{Related Work}
In this section, we introduce some IoT threats, along with a discussion of traditional and LLM-based IoT security-related work, and finally, we present relevant information about CoT.

\subsection{IoT Security Threat and Traditional Methods}
IoT provides ubiquitous sensing and computational capabilities \cite{nguyen2021federated}, having found extensive applications across various domains, including healthcare \cite{balasundaram2023internet}, transportation \cite{song2020applications}, and industrial systems \cite{peter2023industrial}. It is estimated that the global IoT device population is projected to reach 125 billion by 2030 \cite{khan2021federated}. However, the exponential growth of IoT devices, coupled with insufficient security measures, leaves IoT systems highly vulnerable to cyberattacks \cite{schiller2022landscape}, such as botnets \cite{kolias2017ddos}, ransomware \cite{yaqoob2017rise}, advanced persistent threats (APT) \cite{chen2022machine}, and man-in-the-middle (MITM) \cite{salem2021man}. Furthermore, the OWASP Top Ten IoT Security Risks report \cite{ferrara2021static} highlights persistent vulnerabilities in current IoT systems, particularly weak password configurations, firmware vulnerabilities, and unauthorized data access.

To enhance IoT security, researchers have developed various vulnerability detection tools \cite{shoshitaishvili2015firmalice, corteggiani2018inception, zheng2019firm}. Firmalice \cite{shoshitaishvili2015firmalice} automates vulnerability identification through an input determinism model. However, its detection capability is limited in scope and ineffective against complex attack scenarios. Traditional approaches targeting embedded systems and their firmware in IoT have increasingly struggled to address the growing sophistication of security threats. Machine learning (ML) and deep learning (DL), with their capacity to process large and diverse datasets and autonomously learn vulnerability patterns, have attracted growing research interest in IoT security protection \cite{al2020survey, hussain2020machine, shafiq2020corrauc, vasan2020mthael, chaganti2022deep}. For instance, Shafiq et al. \cite{shafiq2020corrauc} proposed CorrAUC, an ML-based method for detecting malicious IoT traffic. It utilizes AUC metrics to filter relevant features and further applies an ensemble of TOPSIS and Shannon entropy to validate the selected features. According to DL, Vasan et al. \cite{vasan2020mthael} introduced the MTHAE model, which combines information gain and OpCode dictionary techniques with hybrid feature selection architectures for IoT malware detection. Despite their advantages in data-driven modeling and automated feature extraction, ML/DL methods still face limitations. Their heavy dependence on data quantity, along with insufficient semantic understanding and contextual reasoning, hinders their ability to provide comprehensive and actionable insights for securing IoT systems.

\subsection{LLM-based Methods for Enhancing IoT Security}
LLMs, as advanced neural network architectures, have achieved remarkable breakthroughs in natural language processing. Typically pre-trained on large-scale multimodal datasets comprising text, code, and other data types, LLMs exhibit strong contextual reasoning capabilities and high-quality decision-making performance \cite{kok2024iot}. Representative models include the Generative Pre-trained Transformer (GPT) series, BERT \cite{devlin2019bert}, LLaMA \cite{touvron2023llama}, and DeepSeek \cite{liu2024deepseek}. Leveraging these advantages, LLMs offer novel technical solutions to address key challenges inherent in IoT, such as data heterogeneity and stringent real-time processing requirements. Through powerful semantic modeling and inference capabilities, LLMs can significantly enhance the understanding and processing efficiency of data streams within IoT systems. Furthermore, the integration of LLMs into resource-constrained environments enables context-aware edge intelligence, allowing IoT systems to better adapt to dynamically changing conditions and improving overall responsiveness and intelligence \cite{aung2025generative}.

As the convergence of LLMs and IoT continues to deepen, increasing research attention has been directed toward their potential in enhancing IoT security tasks, particularly in areas such as vulnerability detection \cite{wang2024llmif, meng2024large, ma2024one}, intrusion prevention \cite{ferrag2024revolutionizing, li2024ids, nwafor2024evaluating}, and threat intelligence analysis \cite{hasan2024distributed, hu2024llm}. For vulnerability detection, Ma et al. \cite{ma2024one} proposed mGPTFuzz, the first fuzzing framework specifically designed for vulnerability discovery in Matter-based IoT devices. This tool leverages LLMs to translate human-readable content from the Matter specification into machine-interpretable representations, specifically as finite state machines. Based on a controller-oriented architecture and tailored fuzzing strategies, mGPTFuzz performs black-box fuzz testing on Matter devices to identify potential security vulnerabilities. For intrusion prevention, Ferrag et al. \cite{ferrag2024revolutionizing} developed SecurityBERT, a network threat detection framework specialized for IoT environments. The system extracts features from network traffic data and employs a BERT-based architecture for model training and fine-tuning. For threat intelligence analysis, Hu et al. \cite{hu2024llm} devised a method for constructing knowledge graphs from unstructured threat intelligence. By leveraging the few-shot learning capabilities of GPT, the framework performs data annotation and augmentation to build a fine-tuning dataset, which is then used to enable automated analysis of textual threat intelligence through the fine-tuned model.

\subsection{Chain-of-Thought}
CoT is a reasoning paradigm that enhances the reasoning capabilities of LLMs by prompting them to generate intermediate reasoning steps before arriving at a final answer \cite{wei2022chain}. Unlike traditional approaches that directly respond to a task, CoT emphasizes a progressive process by decomposing complex problems into a sequence of logically connected sub-problems, thereby mimicking human reasoning. This approach not only improves accuracy in tasks involving arithmetic or logical reasoning but also ensures enhanced interpretability of the decision-making process.

Building upon CoT, Yao et al. \cite{yao2023tree} introduced the Tree-of-Thought (ToT) method, which leverages tree search to expand the reasoning space and discover more optimal reasoning paths that may be overlooked by CoT. However, this method incurs significant additional inference costs. To mitigate this limitation, Zhang et al. \cite{zhang2024chain} further proposed the Chain-of-Preference Optimization (CPO). It utilizes the tree structures generated by ToT to collect preference data at each reasoning step and applies the Direct Preference Optimization (DPO) algorithm to fine-tune the LLM. As a result, the model is guided to select superior reasoning paths while effectively avoiding the high computational overhead associated with ToT.

\section{IoT security assistant}
The increasing complexity and diversity of IoT systems make protecting these environments more challenging than ever before.  IoT devices, ranging from smart appliances to industrial sensors, often operate in heterogeneous environments with varying security requirements. Traditional security solutions struggle to adapt to these dynamic systems, making the need for smarter and more adaptable security assistants even more apparent.

To address this issue, we propose an IoT security assistant driven by LLMs, enhanced through the ICoT process. In this section, we first introduce the design goals of the IoT security assistant, followed by an overview of the basic system architecture.

\subsection{Design Goals}
The design goals of our IoT security assistant can be broken down into the following subgoals:

\begin{itemize}
    \item[1)] \textit{Improved Understanding of IoT Security Threats:} The primary goal is to enhance the LLM's ability to comprehend the nuances of IoT security vulnerabilities and threats. Traditional security models often lack the flexibility to address the diverse range of security issues faced by IoT devices. By employing ICoT, our system can reason step-by-step through security scenarios, allowing for a more comprehensive understanding of potential vulnerabilities.
    \item[2)] \textit{Context-Aware Security Recommendations:} Different IoT environments require different security measures. Our system aims to generate security recommendations that are not only accurate, but also tailored to the specific needs of the user and the IoT system in question. Specifically, we generate correct and personalized security recommendations by breaking down the characteristics of security vulnerabilities and the features of the user.
    \item[3)] \textit{Scalability and Flexibility:} IoT environments are constantly evolving, with new devices and technologies added regularly. Therefore, our security assistant is designed to scale and adapt to new IoT systems without requiring extensive retraining or task-specific fine-tuning. The ICoT method ensures that the assistant can handle various security challenges of the IoT in different contexts.
    \item[4)] \textit{Ease of Use and Integration:} To ensure broad adoption and utility, the system is designed to be user-friendly and easy to integrate into existing IoT networks. The security assistant should provide clear, actionable advice that can be understood by users with varying levels of expertise.
\end{itemize}

\begin{figure}[tp]
    \centering
    \includegraphics[width=\linewidth]{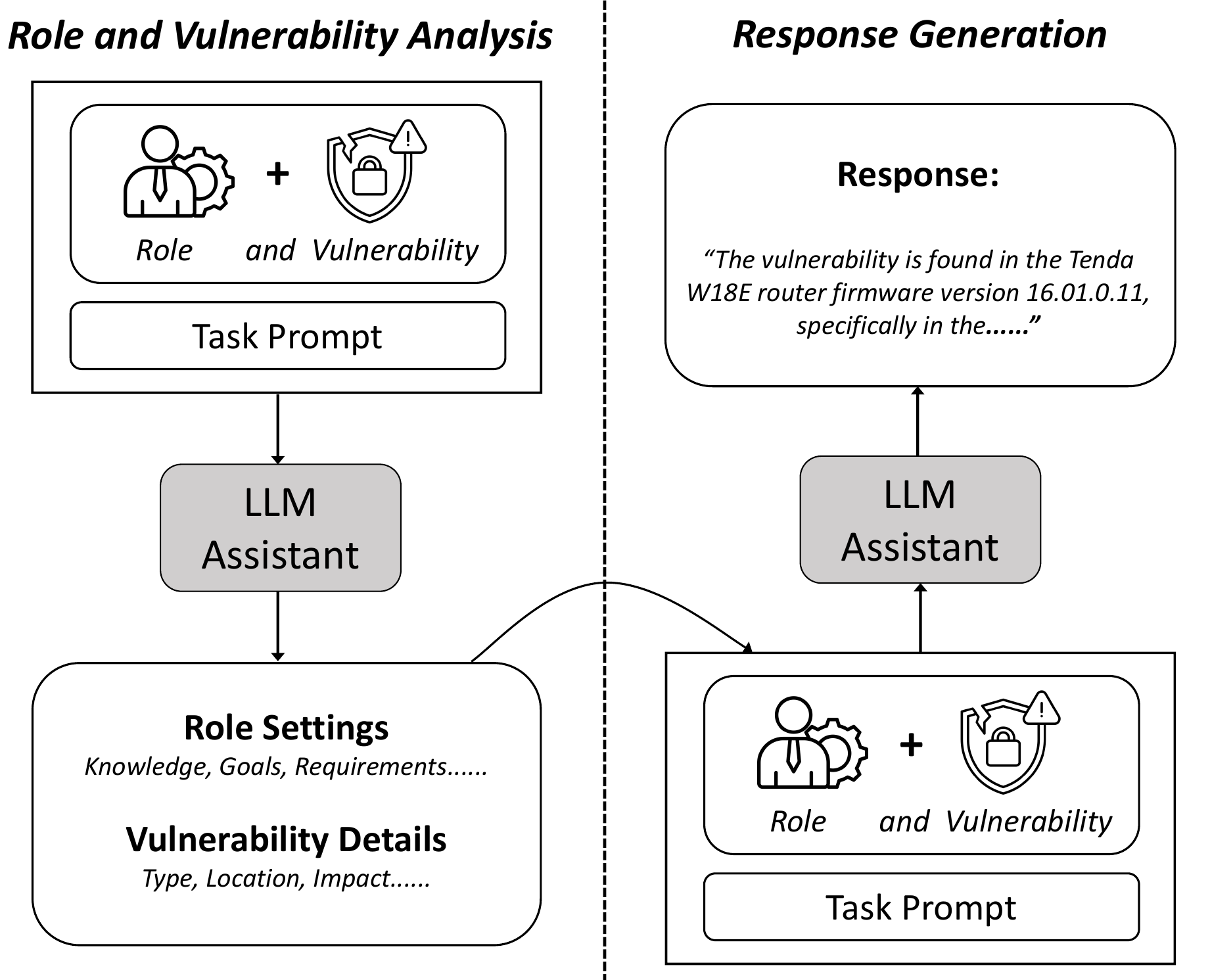}
    \caption{System architecture.} 
    \label{overview}
\end{figure}

\subsection{System Architecture}
The general workflow of the IoT security assistant is shown in Fig.~\ref{overview}. The LLM plays two roles in this process. First, it analyzes the user’s characteristics based on the initial input, such as the user’s knowledge level, goals, and requirements, while also analyzing the characteristics of the vulnerability, clarifying the type, location, and impact of the vulnerability. Second, it serves as the model that generates human-readable security recommendations. After the initial processing, the LLM, with the restructured prompt, generates context-aware and actionable advice based on the professional background and specific needs of the user. The LLM is seamlessly integrated with the CoT engine, ensuring that the reasoning process produces relevant and accurate security solutions.

Specifically, ICoT first conducts a detailed analysis of the role and vulnerability inputs and outputs certain feature values. During the second inference of the model, the first output is used as part of the input to assist in the reasoning. This enables the system to consider factors such as device interconnectivity, communication protocols, and operational environments, allowing for reasoning specific to IoT security scenarios. This module ensures that the assistant’s reasoning is both relevant and effective within the context of IoT systems.

Together, the LLM and ICoT form a powerful, intelligent IoT security assistant capable of providing effective security support in various IoT environments. The system is adaptive and scalable, requiring no specific models or fine-tuning, ensuring its effectiveness in the dynamic and constantly evolving IoT ecosystem.

\section{IoT Chain-of-Thought}

\begin{figure*}[tp]
    \centering
    \includegraphics[width=\linewidth]{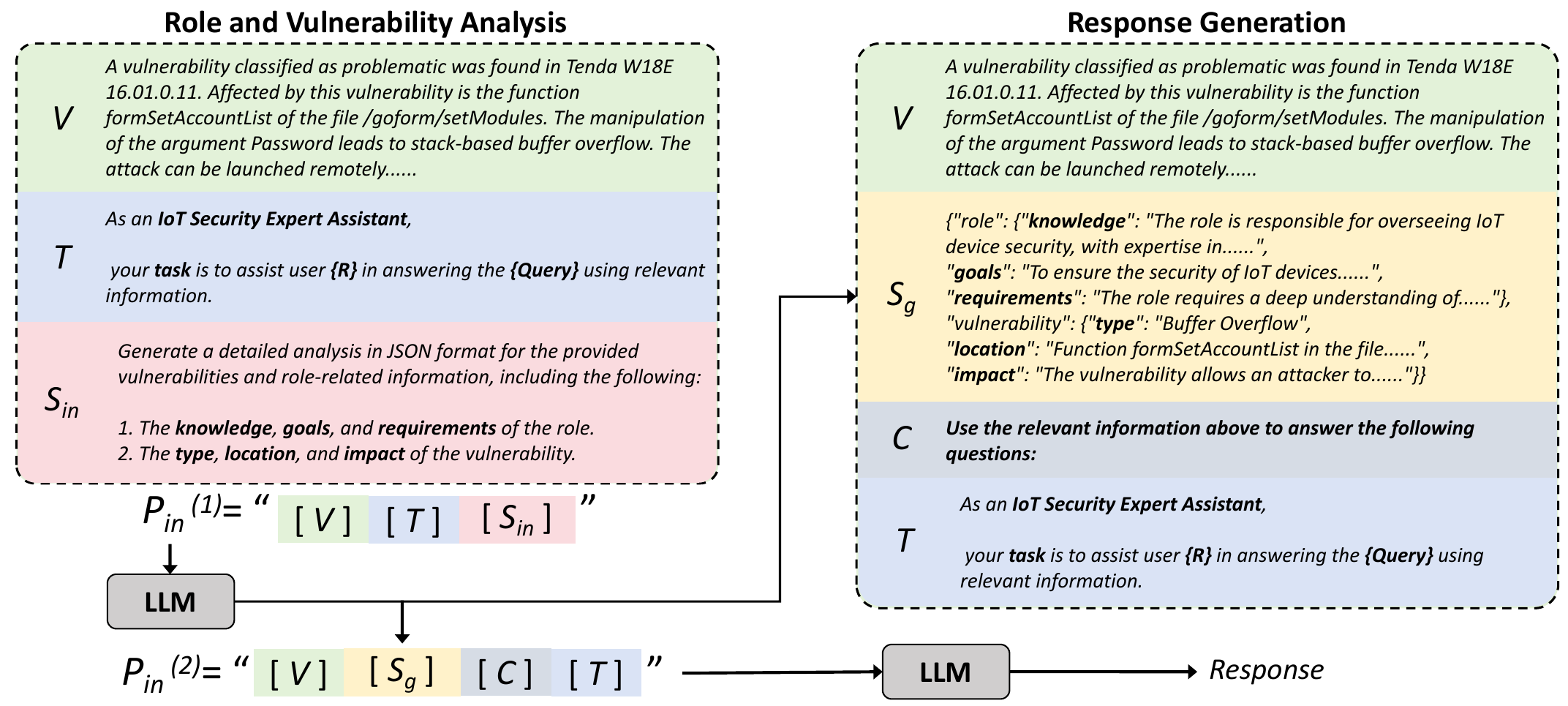}
    \caption{Full prompt example of ICoT.}
    \label{ICoT}
\end{figure*}

In this section, we introduce ICoT, which enhances the capabilities of LLMs to analyze, understand, and mitigate vulnerabilities in IoT devices. By leveraging a systematic reasoning process, ICoT helps LLMs produce more accurate and context-aware security assessments and recommendations for IoT vulnerabilities.

\subsection{Preliminaries}
The LLM is an advanced and highly sophisticated model specifically designed to process textual input and generate coherent, contextually appropriate responses. When presented with an input prompt $P_{in}$, which could be in the form of a question, statement, or instruction, the LLM systematically processes this text through multiple transformer-based layers to produce an output response $R$. This process involves complex interactions between layers that help the model understand and generate human-like text.

Formally, the LLM can be represented as a mathematical function $f_\theta(\cdot)$, where $\theta$ represents the parameters that define the model’s architecture. This function maps the given input prompt to a corresponding output response, as illustrated in the following equation:
\begin{equation} 
R = f_\theta(P_{in}) 
\end{equation}

In this context, the input prompt $P_{in}$ is tokenized, meaning it is converted into smaller, discrete units, and then embedded into a high-dimensional space. This embedding allows the model to capture both the semantic and syntactic relationships present in the text, enabling it to understand the meaning and structure of the input.  The architecture of the model and the pre-training methods used to train the parameters $\theta$ may differ across various LLM implementations. However, the core framework and the fundamental approach remain consistent, with a focus on processing and generating natural language.

Our proposed method, ICoT, introduces a novel approach that does not require any fine-tuning or direct access to the model's underlying parameters $\theta$. Instead, ICoT leverages the inherent reasoning capabilities that are already built into the LLM by using carefully crafted prompts. This makes ICoT a practical and highly efficient technique, as it allows for optimal utilization of the model's pre-existing reasoning power without the need for additional adjustments or retraining.

\subsection{Role and Vulnerability Analysis}
As shown in Fig.~\ref{ICoT}, ICoT integrates the CoT framework into the IoT security context by applying a step-by-step reasoning approach to vulnerability analysis. This method ensures that the model considers a variety of factors related to the IoT device, such as its role in the network, the nature and severity of the vulnerability, and the potential consequences of an attack. By analyzing the vulnerability in this comprehensive manner, the system can generate more reliable, context-aware security advice that helps users address potential risks effectively.

The first step in the ICoT process involves analyzing both the background of the user, $R$, and the characteristics of the vulnerability. This dual analysis ensures that the system's response is accurate, relevant, and specifically tailored to the needs and expertise of the user. This process can be formalized~as:
\begin{equation}
P_{in}^{(1)} =  `` [ V ] [ T ] [ S_{in} ] " 
\end{equation}
where $P_{in}^{(1)}$ represents the input for this iteration, $V$ is the vulnerability description, $T$ is the user’s role and query, and $S_{in}$ is the analysis prompt.

For example, the following is a description of a vulnerability: ``\textit{A vulnerability classified as problematic was found in Tenda W18E 16.01.0.11. Affected by this vulnerability is the function formSetAccountList in the file /goform/setModules. The manipulation of the argument Password leads to a stack-based buffer overflow. The attack can be launched remotely. The exploit has been disclosed to the public and may be used. Shenzhen Tenda Technology Co., Ltd. of W18E has an out-of-bounds write vulnerability in the firmware. Service operation interruption (DoS) may occur.}''

ICoT analyzes this vulnerability by considering the following factors:

\begin{itemize} 
    \item \textit{Vulnerability Type:} In this case, the vulnerability is identified as a buffer overflow, a common and critical issue in IoT systems. 
    \item \textit{Location:} The vulnerability is found within a specific function in the device's firmware, which helps to pinpoint where the issue lies. 
    \item \textit{Impact:} The remote execution of arbitrary code or a denial of service (DoS) attack due to a system crash may result from exploiting this vulnerability. 
\end{itemize}

In this process, user analysis also plays a crucial role alongside device and vulnerability analysis. The system must evaluate the user’s role and level of expertise to generate responses that are not only technically accurate but also comprehensible. This requires consideration of the user’s specific query $Query$. Important aspects of user analysis include:

\begin{itemize} 
\item \textit{User Knowledge:} The system assesses the user’s familiarity with IoT security issues. For instance, a network administrator might require in-depth technical details, whereas a general user may only need basic, actionable steps to mitigate risks. 
\item \textit{User Goals:} The assistant takes into account the user's primary objectives, such as securing a network, preventing unauthorized access, or minimizing system downtime. Understanding these goals helps the system prioritize the most important aspects of the vulnerability that need to be addressed. 
\item \textit{User Requirements:} The system also evaluates the user's practical constraints, such as available technical resources for patching devices, whether they are working in a critical production environment where downtime is unacceptable, or if a temporary solution is needed before a permanent fix can be applied. This part is specifically determined by $Query$.
\end{itemize}

Certainly, the analysis of vulnerabilities and users is not limited to the aspects mentioned above. Given the variety of vulnerabilities found in IoT systems, in addition to the type, location, and impact, other characteristics can also be considered for analysis. For example, factors such as the severity of the vulnerability, the attack vector (e.g., remote or local exploitation), and the likelihood of exploitation based on known threat intelligence can all provide additional insights. Additionally, vulnerabilities may have different levels of exploitability depending on the environment in which the device operates, such as whether it is in a private network or exposed to the public internet.

User analysis follows a similar approach. Apart from understanding the user's technical expertise and goals, more granular user characteristics can be considered. For instance, the user's level of familiarity with specific IoT devices, their previous experience with security incidents, or even their role within an organization can all influence how the system generates a response. In certain cases, users might even define role templates in advance, allowing them to directly input user characteristics into the second phase of the process. This approach eliminates the need for the initial analysis round, making the system more efficient and reducing the processing time required to generate a response. While the current approach involves an initial round of vulnerability and user analysis, ICoT is flexible enough to accommodate a more streamlined process where predefined user profiles are directly input into the response generation phase. This flexibility allows for faster, more targeted recommendations.

The detailed analysis of both the vulnerability and the user is essential for the next step in the ICoT process. The system generates a tailored response based on both the technical details of the vulnerability and the user’s specific context, expertise, and requirements. By incorporating this dual analysis approach, ICoT ensures that the response is not only technically precise but also actionable and relevant to the user's particular situation. Consequently, the LLM generates a JSON format $S_{g}$ that encapsulates both user and vulnerability characteristics, as follows:
\begin{equation} 
S_{g} = f_\theta(P_{in}^{(1)}) 
\end{equation}

\subsection{Response Generation}
Once the role and vulnerability have been thoroughly analyzed, the next phase of the ICoT process focuses on generating specific responses. The input for this phase, $P_{in}^{(2)}$, can be formally represented as:

\begin{equation} 
P_{in}^{(2)} = `` [ V ] [ S_{g} ] [ C ] [ T ]"
\end{equation}

Here, $S_{g}$ represents the output from the first phase, which includes the results of both the vulnerability and user analysis. The $C$ component provides a brief instruction to the LLM, prompting it to consider the provided context. This is explicitly given as: ``\textit{Use the relevant information above to answer the following questions:}''. This ensures that the model integrates all relevant data when generating its response.

This phase is pivotal because it bridges the reasoning process and translates it into practical, actionable advice for the user. By leveraging the insights from the vulnerability and user analysis, the system can now generate customized recommendations. These responses include:

\begin{itemize} 
\item[1)] \textit{Actionable recommendations:} These responses could suggest specific steps, such as patching the identified vulnerability, disabling the affected features, or implementing compensatory security measures to mitigate the risk. The goal is to provide immediate and practical actions that the user can take to address the vulnerability. 
\item[2)] \textit{Security best practices:} Beyond addressing the current vulnerability, the system can also recommend general security practices to help prevent future risks. These might include strategies such as ensuring proper user input sanitization, validating data rigorously, or employing encryption to secure sensitive communications. 
\item[3)] \textit{Context-aware responses:} The advice given will vary depending on the user's role and technical expertise. For a general user, the response may be simpler, recommending easy-to-follow actions like updating firmware or changing passwords. For a developer or network administrator, the suggestions might include more technical measures such as deploying specific patches or modifying the system's code to prevent the vulnerability from reappearing. 
\end{itemize}

By incorporating role-specific context and a detailed understanding of security threats, the system ensures that the advice it provides is both precise and highly relevant to the user's situation. This context-aware approach guarantees that the responses are not only actionable but also appropriately tailored to the user's capabilities and needs. Thus, the LLM generates the final response $R$ for the vulnerability, user, and prompt, as follows:

\begin{equation} 
R = f_\theta(P_{in}^{(2)}) 
\end{equation}

\section{Experiments}
We applied the ICoT method to four popular LLMs: GPT-4o~\cite{achiam2023gpt}, GPT-4o-mini, DeepSeek-R1~\cite{guo2025deepseek}, and DeepSeek-V3~\cite{liu2024deepseek}. Our primary goal is to determine whether ICoT can effectively generalize and enhance the capabilities of state-of-the-art LLMs in handling IoT security issues.

\subsection{Setup}
\textbf{Dataset.} We have gathered two types of datasets related to IoT security and threats. The first, VARIoT Vulnerabilities~\cite{janiszewski2021automatic}, compiles a collection of known vulnerabilities in IoT devices, along with detailed descriptions of the associated risks. The second, VARIoT Exploits~\cite{janiszewski2021automatic}, focuses on exploitations targeting IoT devices, providing insights into these vulnerabilities from the perspective of attackers.

\textbf{Model.} We connect four popular LLMs via APIs to serve as the foundation for our IoT security assistant:
\begin{itemize} 
    \item[1)] \textit{GPT-4o:} Developed by OpenAI, GPT-4o is a multimodal model that excels in natural language understanding and generation. It supports text, image, and audio inputs, making it suitable for diverse applications such as voice assistants and real-time translation.
    \item[2)] \textit{GPT-4o-mini:} GPT-4o-mini is a more compact and cost-effective version of GPT-4o. It offers a balance between performance and efficiency, making it ideal for integration into services with high API call volumes.
    \item[3)] \textit{DeepSeek-R1:} A Chinese-developed model by DeepSeek, DeepSeek-R1 is optimized for reasoning tasks such as mathematics, code generation, and logical inference. It employs reinforcement learning techniques to enhance its reasoning capabilities.
    \item[4)] \textit{DeepSeek-V3:} DeepSeek-V3 is a general-purpose model that performs well across various natural language processing tasks. It utilizes a mixture-of-experts architecture to balance performance and computational efficiency.
\end{itemize}

\subsection{Implementation Details}
Although there is a substantial amount of research on IoT security, there is still a lack of publicly available labeled datasets for IoT security vulnerabilities and threat analysis. To evaluate the effectiveness of our ICoT method, we compared the outputs generated by ICoT with those produced by an LLM that does not include the IoT-specific reasoning steps provided by our Chain-of-Thought approach. To assess the quality of the generated answers, we used an independent LLM as an evaluator (fixed as GPT-4o) and measured the results based on five metrics: Accuracy, Relevance, Detail, Technicality, and Friendliness, which are defined as follows:

\begin{itemize} 
    \item[1)] \textit{Accuracy:} The correctness of the answer, ensuring that it aligns with established IoT security principles and accurately addresses the described vulnerabilities or threats.
    \item[2)] \textit{Relevance:} The extent to which the answer directly addresses the specific question posed by the IoT security scenario and meets the user's needs, considering their professional backgrounds and context.
    \item[3)] \textit{Detail:} The level of detail provided in the answer, including how well the answer elaborates on the issue and offers a thorough explanation of the security implications.
    \item[4)] \textit{Technicality:} The response should demonstrate a deep understanding of IoT technologies and their security implications, particularly in relation to IoT protocols, standards, and security measures, ensuring the accuracy and appropriateness of the technical language used.
    \item[5)] \textit{Friendliness:} The ease of understanding of the answer and its practical value to the user. This includes how well the response translates technical information into actionable security steps or solutions that are tailored to the user’s personalized context.
\end{itemize}

\begin{figure}[tp]
    \centering
    \includegraphics[width=.65\linewidth]{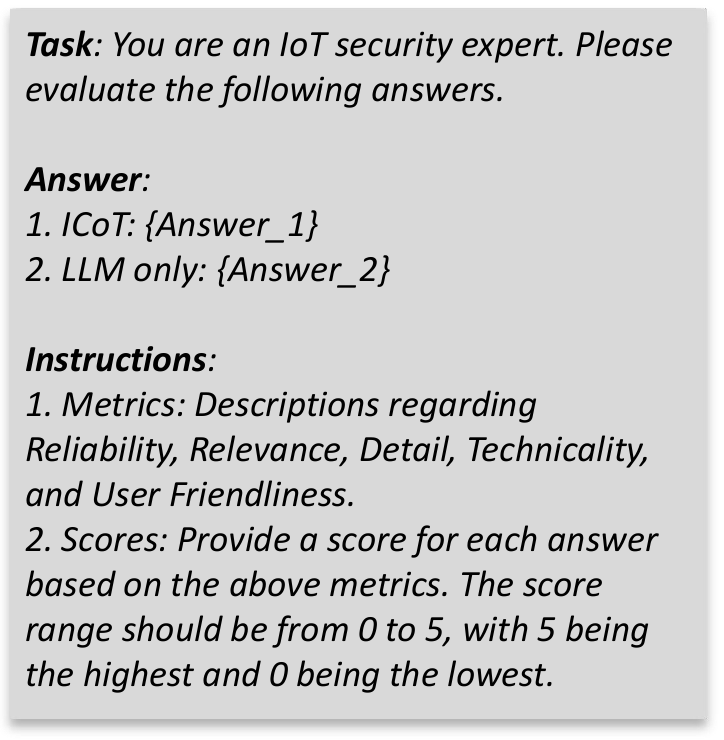}
    \caption{The evaluation template.} 
    \label{eva}
\end{figure}

The scores for all five metrics are fixed within the range of [0, 5], with 5 representing the highest quality. The evaluator receives the answers from both ICoT and the LLM-only approach at the same time, ensuring a uniform assessment of both response sets. This approach minimizes the impact of any inherent randomness in the LLM outputs, allowing for a fair and unbiased comparison between the two models~\cite{dong2025chatiot}. The evaluation framework is shown in Fig.~\ref{eva}.

\subsection{Main Results}
As shown in Table~\ref{Result}, ICoT enhances the performance of LLMs in the field of IoT security. ICoT demonstrates notable improvements across three user roles: General User, Developer, and Technical Officer. Moreover, the personalized responses generated by ICoT lead to increased user friendliness and relevance of the recommendations. However, not all improvements are significant. In addition to the inherent nature of the vulnerability descriptions themselves, one possible reason for this is that ICoT sometimes encourages the LLM to generate excessive redundant information, which may diminish the overall clarity or conciseness of the responses.

\begin{table*}
    \centering
    \caption{Comparison of ICoT with LLM only method.}
    \label{Result}
    \resizebox{\textwidth}{!}{
    \begin{threeparttable}
        \begin{tabular}{c|c|cc|cc|cc|cc}
        \toprule
          \multirow{2}*{Role} & \multirow{2}*{Metric} &\multicolumn{2}{c|}{GPT-4o} & \multicolumn{2}{c|}{GPT-4o-mini}&\multicolumn{2}{c|}{DeepSeek-V3}&\multicolumn{2}{c}{DeepSeek-R1}\\ 
          \hhline{~~--------}
          ~ & ~ & \textbf{ICoT} & LLM only &  \textbf{ICoT} & LLM only & \textbf{ICoT} & LLM only & \textbf{ICoT} &  LLM only \\ 
          \hline
          \multirow{5}*{General User}  & Reliability & 4.61(+0.08) & 4.53 & 4.67(+0.39) & 4.28 & 4.77(+0.05) & 4.72 & 4.42(+0.47) & 3.95\\ 
          ~ &  Relevance & 4.20(+0.14) & 4.06 & 4.34(+0.84) & 3.50 & 4.98(+0.99) & 3.99 & 4.94(+0.14) & 4.80\\
          ~ &  Detail & 4.37(+0.37) & 4.00 & 4.20(+0.10) & 4.10 & 4.81(+0.79) & 4.02 & 4.31(+0.48) & 3.83\\
          ~ & Technicality & 4.47(+0.81) & 3.66 & 4.43(+0.88) & 3.55 & 4.96(+0.25) & 4.71 & 4.65(+0.31) & 4.34\\
          ~ & Friendliness & 4.72(+0.09) & 4.63 & 4.04(+0.60) & 3.44 & 4.93(+0.55) & 4.38 & 4.77(+0.80) & 3.97\\
          \hline

          \multirow{5}*{Developer} &  Reliability & 4.27(+0.74) & 3.53 & 4.50(+0.90) & 3.60 & 4.75(+0.62) & 4.13 & 4.68(+0.93) & 3.75\\
          ~  & Relevance & 4.36(+0.66) & 3.70 & 4.71(+0.10) & 4.61 & 4.28(+0.61) & 3.67 & 4.03(+0.95) & 3.08\\
          ~  & Detail & 4.35(+0.47) & 3.88 & 4.37(+0.84) & 3.53 & 4.09(+0.74) & 3.35 & 4.43(+0.77) & 3.3.66\\
          ~  & Technicality &  4.80(+0.65) & 4.15 & 4.31(+0.34) & 3.97 & 4.52(+0.31) & 4.21 & 4.41(+0.94) & 3.47\\
          ~ & Friendliness & 4.48(+0.62) & 3.86 & 4.16(+0.99) & 3.17 & 3.86(+0.25) & 3.61 & 4.48(+0.18) & 4.30\\
          \hline

          \multirow{5}*{Technical Officer}&  Reliability & 4.52(+0.08) & 4.44 & 4.23(+0.67) & 3.56 & 4.29(+0.27) & 4.02 & 4.36(+0.21) & 4.15\\
          ~ & Relevance & 4.53(+0.27) & 4.26 & 3.97(+0.40) & 3.57 & 4.43(+0.47) & 3.96 & 4.66(+0.15) & 4.51\\
          ~ & Detail & 4.87(+0.05) & 4.82 & 4.02(+0.34) & 3.68 & 4.65(+0.11) & 4.54 & 4.18(+0.24) & 3.94\\
          ~ & Technicality & 4.06(+0.08) & 3.98 & 4.06(+0.21) & 3.85 & 3.80(+0.73) & 3.07 & 4.69(+0.62) & 4.07\\
          ~ & Friendliness & 4.59(+0.42) & 4.17 & 4.33(+0.11) & 4.22 & 4.99(+0.44) & 4.55 & 4.95(+0.38) & 4.57\\

    \bottomrule
    \end{tabular}
    \end{threeparttable}  
    }
\end{table*}

\subsection{Further Analysis}
In addition to the inherent nature of the vulnerability descriptions themselves, one possible reason for the lack of significant improvement in certain cases is that ICoT sometimes encourages the LLM to generate excessive redundant information.  This redundancy can negatively impact the clarity and conciseness of the responses, making them less efficient and harder for the user to interpret.  While the addition of more context might seem beneficial, it can overwhelm the user with unnecessary details, particularly when concise and actionable advice is needed.

Furthermore, LLMs, despite their powerful reasoning capabilities, are not immune to hallucination—a phenomenon where the model generates information that is plausible-sounding but fabricated. In the case of IoT security, hallucinated content could include incorrect security recommendations or unverified details about vulnerabilities.  Such errors can be especially harmful in the context of security analysis, where inaccurate advice could lead to improper mitigation measures or overlooked threats.

While the CoT approach enhances the reasoning process by breaking down problems step by step, it is not always foolproof. CoT is reliant on the model's internal understanding of the problem, which, despite being structured, may still lead to unreliable conclusions in certain situations~\cite{chenreasoning}. The reasoning process may become convoluted or inconsistent when the model struggles with particularly complex or poorly defined vulnerabilities, leading to incorrect or incomplete security assessments.

The presence of redundant information, hallucinations, and inconsistencies in reasoning illustrates the ongoing challenges in applying LLMs to real-world security scenarios. Tackling these issues is essential for enhancing the reliability and performance of ICoT and similar systems in the realm of IoT security.

\subsection{Scalability}
Although only three user roles are used in the experiment, we do not restrict the specific identity of the user in practice. Users can even import their own identity templates (during the second input). Similarly, we do not set a fixed number of characteristics to analyze the vulnerabilities; users can adjust this themselves. From various perspectives, ICoT offers a high level of scalability.

\section{Discussion}
In this section, we discuss the implications, challenges, and future directions of the ICoT method in the context of IoT security. The ICoT method represents a significant step forward in understanding and addressing IoT security vulnerabilities, but several aspects warrant further exploration.

\subsection{Impact on the IoT Security Field}
With the widespread adoption of IoT devices, the security challenges have become increasingly significant, posing major threats to global cybersecurity. The ICoT method proposed in this paper enhances the performance of LLMs in analyzing IoT security vulnerabilities by introducing the CoT reasoning mechanism. Compared to traditional vulnerability detection and response methods, ICoT provides more accurate and in-depth security analysis, especially in handling complex vulnerabilities and dynamic security threats.

The impact of ICoT on the IoT security field can be summarized in several key aspects:

\begin{itemize} 
    \item[1)] \textit{Improved Vulnerability Identification Accuracy:} Traditional vulnerability detection tools often rely on static rules or pattern matching, which can miss emerging vulnerabilities or complex attack patterns. The ICoT method, through step-by-step reasoning, provides a more comprehensive identification of vulnerabilities and potential risks, leading to more reliable security assessments. 
    \item[2)] \textit{Automated Security Response:} ICoT can generate customized security recommendations based on different devices' roles and vulnerability types. This automation reduces the need for manual intervention and significantly enhances the efficiency of security management. 
    \item[3)] \textit{Cross-Domain Application Potential:} While the focus of this paper is IoT security, the ICoT method can also be extended to other security domains, such as Industrial Control Systems (ICS), smart grids, and smart home security. 
\end{itemize}

\subsection{Challenges and Limitations}
Although the ICoT method performs well both theoretically and in experiments, it still faces several challenges and limitations in practical applications:

\begin{itemize} 
    \item[1)] \textit{Diversity and Heterogeneity of IoT Devices:} IoT devices come in many varieties, each with different hardware architectures, firmware versions, and security configurations. The reasoning process in ICoT needs to be adapted for different devices, placing high demands on the model's generalization ability. Some devices may lack sufficient security information or fail to release timely patches, making vulnerability analysis more difficult. 
    \item[2)] \textit{Incomplete Vulnerability Data:} IoT device security vulnerability information is often scattered and updated slowly, meaning ICoT may not have access to the latest vulnerability data. Additionally, descriptions of vulnerabilities and their remediation methods may differ between manufacturers, which can make the model less effective in addressing new vulnerabilities. 
    \item[3)] \textit{Inference Efficiency and Real-Time Performance:} While ICoT provides powerful reasoning capabilities, large-scale vulnerability detection and inference tasks can impose significant computational demands. Especially in resource-constrained IoT environments, the efficiency and response time of the inference process could become a performance bottleneck. 
    \item[4)] \textit{Security and Privacy Concerns:} While ICoT can improve security, the large-scale deployment of such systems may involve user privacy and data security issues. In particular, when dealing with sensitive data, ensuring the privacy and security of information remains a challenge that needs to be addressed~\cite{dhinakaran2024privacy,magara2024internet}. 
\end{itemize}

\subsection{Future Work}
The ICoT method provides a new framework for IoT security analysis, but many avenues for further research and improvement remain:

\begin{itemize} 
    \item[1)] Currently, ICoT is mainly applied to common IoT devices and protocols. Future research can extend ICoT to include more types of IoT devices and communication protocols, such as Low Power Wide Area Network (LPWAN) devices, Industrial IoT (IIoT) devices, etc., to improve its applicability and generalizability. 
    \item[2)] To address the rapid changes in IoT environments and the emergence of new vulnerabilities, the ICoT system needs to further enhance its adaptability, enabling automatic adjustment of reasoning strategies and response plans to deal with new security threats. This may involve online learning or incremental training to continuously optimize the model. 
    \item[3)] Future work can explore integrating more types of sensor data, network traffic information, and system logs into ICoT, further improving its ability to recognize and respond to complex security scenarios. For example, combining visual, audio, and environmental data can enhance the model's capacity for multimodal security analysis. 
    \item[4)] To accommodate large-scale IoT networks, ICoT needs to further optimize its inference efficiency, especially in resource-constrained IoT devices. Research into reducing computational and storage requirements while maintaining high accuracy will be key to enhancing the application potential of this technology. 
    \item[5)] With the widespread deployment of ICoT, safeguarding data security and privacy will be an important research direction. Developing privacy-preserving inference mechanisms and ensuring that IoT security assistants do not leak user privacy while processing sensitive data will help increase user trust in this technology. 
\end{itemize}
 
\section{Conclusion}
In this paper, we introduced an IoT security assistant powered by LLM and enhanced through the ICoT process. Through the ICoT we proposed, the LLM is better equipped to understand and address the complexity of IoT security vulnerabilities, providing more accurate, context-aware security recommendations that align with the user’s personalized needs. Specifically, ICoT breaks down the vulnerability and user characteristics through a two-stage analysis to generate more effective security advice. Experimental results show that our approach significantly enhances the ability to analyze vulnerabilities and mitigate security threats, without requiring additional training or fine-tuning. Our system not only improves the LLM's capabilities in the field of IoT security, but also offers a new framework for integrating structured reasoning into large-scale security applications. Future work will explore further optimization of the ICoT process and expand the system’s application to a broader range of IoT security challenges, ultimately providing stronger and more efficient security solutions for the rapidly growing IoT ecosystem.

\bibliographystyle{IEEEtran}
\bibliography{IEEEabrv,mylib}

\end{document}